\documentclass[12pt]{article}
\input epsf
\newcommand{\email}[1]{\texttt{email: #1}}
\newcommand{\sn}{{\rm sn}} 
\begin{document}
 \title{Density Matrices for a Chain of Oscillators}
 \author{
Ingo Peschel\thanks{\email{peschel@physik.fu-berlin.de}}
\, and Ming-Chiang Chung\thanks{\email{mcchung@physik.fu-berlin.de}} \\
{\small Fachbereich Physik, Freie Universit\"at Berlin,} \\
{\small Arnimallee 14, D-14195 Berlin, Germany}}
 \maketitle
 \begin{abstract}
  We consider chains with an optical phonon spectrum and study the reduced 
  density matrices which occur in density-matrix renormalization group (DMRG)
  calculations. Both for one site and for half of the chain, these are 
  found to be exponentials of bosonic operators. Their spectra, which are 
  correspondingly exponential, are determined and discussed. The results for
  large systems are obtained from the relation to a two-dimensional Gaussian 
  model.
 \end{abstract}
 \section{Introduction}
  The success of the density-matrix renormalization method (DMRG) in treating
  one-dimensional quantum systems \cite{white,DMRG}
  is closely related to the  properties of the involved 
  density matrices. In the procedure, one determines
  the eigenvectors of these matrices and uses those with the largest
  eigenvalues  as a truncated basis. To be able to single out a relatively
  small number, however, the density-matrix spectrum has to decrease 
  rapidly enough. Indeed, it is usually found in the numerical calculations 
   that the eigenvalues decay roughly exponentially.

  In a previous publication \cite{peschk} it was pointed out that, for 
  non-critical integrable models, the exponential behaviour is ultimately 
  a consequence of the Yang-Baxter equations. For two spin one-half models,
  the transverse Ising chain and the uniaxial Heisenberg chain, analytical
  formulae were given and verified in detail in DMRG calculations.

  In the present article, we want to extend these considerations to phonons,
  i.e. to a bosonic problem. So far, comparatively few DMRG studies
  have been dealing with bosons \cite{jzw}-\cite{KWM}. 
  The difference to spin systems is that 
  the full Hilbert space always has infinite dimension. Therefore any 
  numerical treatment has to start with a truncation. One can do this in 
  analogy with the DMRG procedure by selecting local states via the density
  matrix for a single site \cite{zhang,jzw}. This is still a nontrivial 
  quantity with an infinite number of eigenstates in a full treatment, and 
  it is of interest to find its properties in a solvable case. The same
  holds, of course, for the more complicated density matrix of a half-chain
  which is used in the DMRG algorithm.            

  The system we study is a purely bosonic model, a chain of $L$ harmonic 
  oscillators with frequency $\omega_0$, coupled together by springs.
  It has a gap in the phonon spectrum and is a non-critical integrable 
  system as the spin models mentioned above. We write the Hamiltonian
  \begin{equation}
    H\;=\; \sum_{i=1}^{L}\,(-\frac{1}{2}\frac{\partial^2}{\partial x_i^2}\,
    + \frac{1}{2}\,\omega_0^2\, x_i^2\,) \,+\, \sum_{i=1}^{L-1}\,
    \frac{1}{2}\, k \,(x_{i+1}-x_i)^2 \label{eqn:H}
  \end{equation} 
  and will frequently use the form $\omega_0\,=\,1-k$, so that for $k=0$ there
  is no dispersion, while for $k \rightarrow 1$ the spectrum 
  becomes acoustic and the system critical. 
  
  We first consider in Section \ref{sec:one} the density matrix $\rho_1$ 
  for one oscillator and show that it can always be written as the exponential 
  of the Hamiltonian of a (new) harmonic oscillator.
   The spectrum therefore is purely 
  exponential, with a decay rate depending on $k$ and (weakly) on the chosen
  site. This generalizes the known result for the case $L=2$ \cite{Han}.
  The eigenfunctions have the character of squeezed states and
  are used later for numerical calculations.
  In Section \ref{sec:half}, we turn to the density matrix
  $\rho_{h}$  for half
  of the system. We treat the case of small and large $L$ explicitely and 
  find that $\rho_h$ has the same exponential form, with the number of 
  oscillators in the exponent determined by the size of the system.
  The result in the thermodynamic limit is derived by relating the chain
  to a massive two-dimensional Gaussian model 
  and its corner transfer matrices (CTMs).
  It is very similar to that for the spin chains in 
  \cite{peschk} which lead to 
  fermionic operators instead of bosonic ones.
  In particular, the spectrum without the degeneracies
  is purely exponential. Its form for different values of $k$ and different 
  sizes $L$ is discussed in more detail in Section \ref{sec:spectra}, 
  including numerical results obtained by truncation and by DMRG calculations.
  These also illustrate, to what extent the degeneracies are reproduced in an 
  approximate treatment. The concluding Section \ref{sec:concl} contains a 
  summary and additional remarks. Some details concerning the case $L=4$
  and the Gaussian model are given in two appendices.

 \section{Density Matrix for One Oscillator}\label{sec:one}
  In this section we consider the case where one oscillator is
  singled out and all other form the environment. 
  The corresponding density matrix (determined numerically)
  was used previously in the study of an electron-phonon
  system \cite{zhang}. Here, it can be found analytically.
   
  The ground state of $H$ in (\ref{eqn:H}) has the form
  \begin{equation}
    \Psi(\mbox{\boldmath$x$}) \;= \;C\:\cdot\:
     \exp{(-\frac{1}{2}\sum_{ij}\,A_{ij}x_{i}x_{j})}
     \label{eqn:psi}
  \end{equation}
  where $\mbox{\boldmath$x$}\;=\;(x_{1},x_{2},\cdots,x_{L})$. The matrix 
  \begin{equation}
    A_{ij}\;=\;\sum_{q}\,\omega_{q}\phi_{q}(i)\phi_{q}(j)
  \end{equation}  
   is determined by the frequencies $\omega_{q}$ and the eigenvectors 
   $\phi_{q}(i)$ of the normal modes. From the total density matrix
  \begin{equation}
    \rho(\mbox{\boldmath$x$,\boldmath$x'$})\;=\;\Psi(\mbox{\boldmath$x$})\Psi(
    \mbox{\boldmath$x'$})
  \end{equation}  
   one then obtains the reduced one for oscillator $l$ by integrating over all 
   other coordinates $x_{i}\:=\:x'_{i}$. This leads to 
  \begin{equation}
   \rho_{1}(x_{l},x'_{l})\;=\;C_{1}\,\cdot\,
   \exp{(-\frac{1}{2}(a-b)x_{l}^{2}\,)}\,
   \exp{(-\frac{b}{4}(x_{l}-x'_{l})^2\,)}\,
   \exp{(-\frac{1}{2}(a-b) {x'_{l}}^2\,)}
  \label{eqn:D1}
  \end{equation}
  with the constants 
  \begin{eqnarray}
   a\;&=&\; A_{ll} \\
   b\;&=&\; \sum_{i,j\not=l}\:A_{li}\:[A^{(l)}]_{ij}^{-1}\:A_{jl}
  \end{eqnarray}                    
  where $A^{(l)}$ is the matrix obtained from $A$ by deleting the
  $l$-th row and column. The second exponential in (\ref{eqn:D1}) can be 
  transformed into a differential operator, giving
  \begin{equation}
   \rho_{1}\;=\;C_{2}\,\cdot\,\exp{(-\frac{1}{4}\omega^{2}y^{2}\,)}\,
   \exp{(\frac{1}{2}\frac{\partial^2}{\partial {y}^{2}}\,)}
   \,\exp{(-\frac{1}{4}\omega^{2}y^{2}\,)}
  \label{eqn:D2}
  \end{equation} 
  where $y^{2}\,=\,b\,x_{l}^{2}/2$ and $\omega^{2}/4=(a-b)/b$.
  Writing this in terms of Bose operators $\alpha$ , $\alpha^{\dagger}$ 
  one can bring it into diagonal form by an
   equation-of-motion method. The necessary Bogoljubov transformation is 
  \begin{equation}
   \beta\:=\: \cosh\theta\cdot\alpha\:+\sinh\theta\cdot\alpha^{\dagger} 
  \end{equation} 
  with
  \begin{equation}
    e^{\theta} \:=\:(1\,+\,\frac{\omega^2}{4})^{1/4}
  \end{equation}
  As a result, one finds that $\rho_1$ has the form
  \begin{equation}
    \rho_1\:=\:K\,\cdot\,\exp{(-{\cal H}\,)} \label{eqn:calH}
  \end{equation}
   where
  \begin{equation}
    {\cal H}\:=\: \varepsilon\,\beta^{\dagger}\,\beta
  \end{equation}
  is the Hamiltonian of 
  a harmonic oscillator with energy 
  \begin{equation}
    \varepsilon \;=\; 2\sinh^{-1}(\frac{\omega}{2})~~
     \;=\; 2\sinh^{-1}\left(\sqrt{a/b-1}\right) 
  \end{equation}
  Therefore the eigenvalues of $\rho_1$ are $w_n\,=\,K\,
  e^{-\varepsilon n}, n\geq 0$, 
  and the spectrum is purely exponential. The constant
  $K$ follows from the sum rule ${\rm Tr}(\rho_1) = \sum_n w_n =1$. 
  
  \begin{figure}
 \begin{center}
 \epsfxsize=100mm
 \epsfysize=80mm
 \epsffile{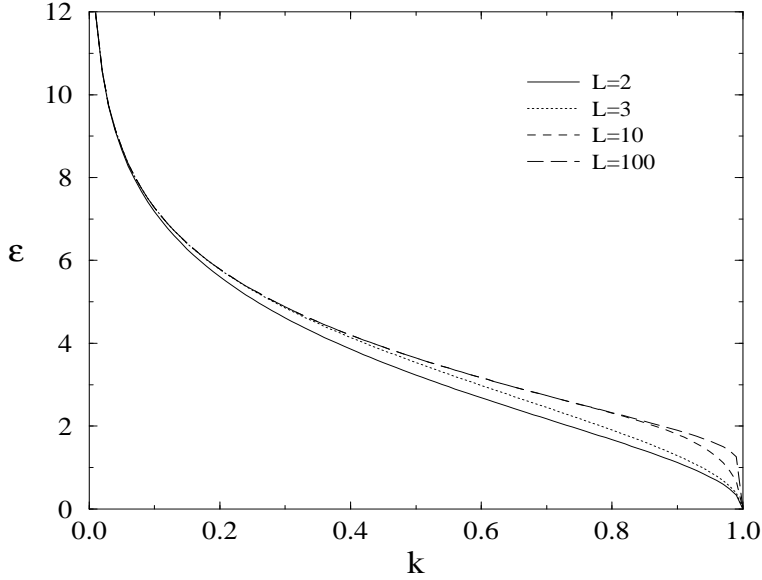}
 \caption{\label{fig1} Eigenvalue $\varepsilon$ in the density matrix for 
  an oscillator at the end of a chain, as a function of $k$
  for different lengths $L$ and $\omega_0=1-k$}
 \end{center}
 \end{figure}

  This result is completely general. The details of the oscillating system and 
  the position of the chosen oscillator enter only via the ratio $a/b$. The 
  same constants $a$ and $b$ also determine the probability to find a certain 
  elongation $x_l$. However, as seen from 
  $\rho_1(x_l,x_l)$ in (\ref{eqn:D1}),
  this quantity depends on the difference ($a-b$) and thus has no direct 
  relation to $\varepsilon$.
    
  In the simplest case of two oscillators ($L\,=\,2$) one finds explicitely 
  \begin{equation}
    \varepsilon\:=\: 2\,\sinh^{-1}\left(\sqrt{4\omega_1\omega_2/
     (\omega_1-\omega_2)^2}\right)
  \end{equation} 
  or, equivalently,
  \begin{equation}
   \varepsilon\:=\: \ln(\coth^2(\frac{\eta}{2}))
  \end{equation}   
  where $\omega_1\,=\,\omega_0$ , $\omega_2\,=\,\sqrt{\omega_0^2+2k}$ are the 
  two eigenfrequencies and $e^{2\eta}\,=\,\omega_2/\omega_1$. This is the 
  result obtained in \cite{Han} in a different way.

 In Fig.\ref{fig1}, $\varepsilon$ is shown as a function of $k$, putting 
  $\omega_0 \,=\,(1-k)$. For $k\rightarrow0$ it diverges 
  logarithmically. In this limit the influence of the second oscillator 
  vanishes, $\Psi(\mbox{\boldmath$x$})$ becomes a product state and one is 
  left with only one nonzero eigenvalue $w_0\,=\,1$.
   For $k\rightarrow1$, on 
  the other hand, $\varepsilon$ goes to zero as $\sqrt{1-k}$ and the
 eigenvalues $w_n$ decrease only very slowly, which reflects the
 strong coupling. These features are encountered also in all other cases.
   For $L=3$
 one can still give explicit analytical formulae, but for larger $L$ the 
 problem has to be treated numerically. 
  In the figure, two additional cases, $L=10$ 
  and $L=100$, are shown. The limit $L\rightarrow\infty$, 
  which is approached exponentially in $L$ with 
  a correlation length increasing with $k$, is indistinguishable
  from $L=100$ on the given scale.

   \begin{figure}
 \begin{center}
 \epsfxsize=100mm
 \epsfysize=80mm
 \epsffile{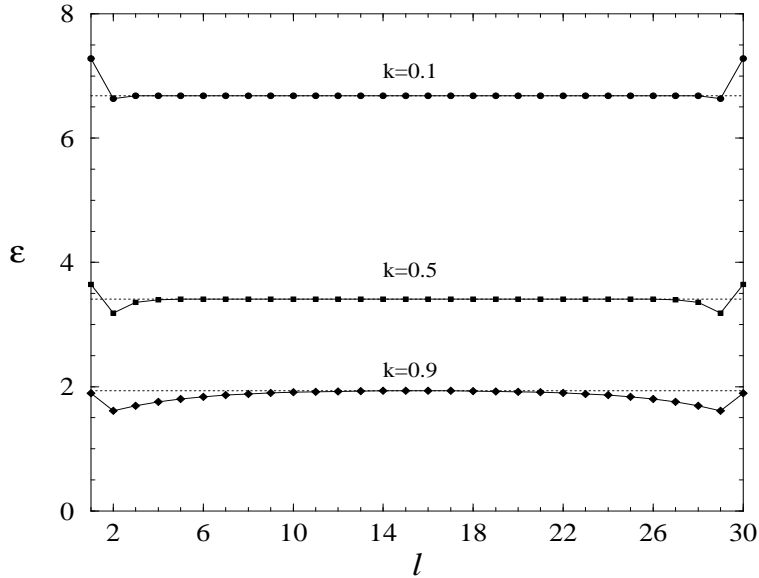}
 \caption{\label{fig2} Eigenvalue $\varepsilon$ as a function of the position
   of the oscillator, for a chain of $L = 30$ sites and three different
   values of $k$.}
 \end{center}
 \end{figure} 

  One can also investigate how $\varepsilon$ varies with the position along
  the chain. The result is shown in Fig.\ref{fig2} for several values of k. 
  One sees that $\varepsilon$ is large at the ends. This corresponds to the 
  fact that the influence of the environment is smaller there. At the next 
  site, however, $\varepsilon$ drops and then approaches
  the bulk value from below as one moves 
  into the interior. The approach becomes slower as
  $k$ increases. The overall differences 
  in the $\varepsilon$-values are not 
  very large, though, as seen in the figure.       
 
  Due to the form of $\rho_1$, its eigenstates are standard oscillator 
  functions of a coordinate $z$ which differs from $x_l$ by 
  a scale factor. Compared with the eigenfunctions of the 
  uncoupled oscillator $l$, they are squeezed states whose 
  spatial extent is reduced by a factor $q=\sqrt{\omega_0/\gamma}$,
  where $\gamma=\sqrt{a(a-b)}$. For small $k$, $q$ approaches one
  and the two sets of functions coincide. 
  With increasing $k$, the amount of squeezing increases, and 
  it is then advantageous to choose the squeezed states as a local basis.
  This was done in the numerical calculations 
  which will be presented in Section \ref{sec:spectra}.

 \section{Density Matrix for a Half-Chain} \label{sec:half}
  We now turn to the central quantity in the usual DMRG calculations, the 
  reduced density matrix for half of the system. It enters each time the 
  system is enlarged in the infinite-size algorithm. We will determine its
  spectrum in the two limits of small and large $L$.

  For $L=2$, one-half of the system is just one oscillator and $\rho_h$ has
  already been obtained in Section \ref{sec:one}. We therefore proceed
  immediately to the case $L=4$. First, we note that the square root
   $\rho_h^{1/2}$ follows
   directly from $\Psi$. If the coordinates along the chain are $(x_2,x_1,
   x'_{1},x'_{2})$, one has
  \begin{equation}
   \rho_h^{1/2}(x_1,x_2;x'_1,x'_2)\;=\; \Psi(x_2,x_1,x'_1,x'_2)
  \end{equation}  
  Taking into account the form (\ref{eqn:psi}) and the symmetries, this 
  leads to
 \begin{equation}
   \rho_h^{1/2}\;=\; C\,\cdot\,
   \exp\left\{-\frac{1}{2}\sum_{ij}a_{ij}(x_{i}x_{j}+
   x'_{i} x'_{j})-\sum_{ij}b_{ij}x_{i}x'_{j}\right\}
   \label{eqn:fourd}
  \end{equation}  
  where the symmetric $(2 \times 2)$ matrices $a_{ij}$ and $b_{ij}$
  follow from the matrix $A$ of Section \ref{sec:one}.
  Altogether one has six different coefficients which couple the
  variables as shown in the following figure.
     
  \unitlength1cm
  \begin{center}
  \begin{picture}(5,5)
  \put(1,1){\line(1,0){3}}
  \put(1,1){\line(0,1){3}}
  \put(1,4){\line(1,0){3}}
  \put(4,1){\line(0,1){3}}
  \multiput(1,1)(0.1,0.1){30}{\circle*{0.01}}
  \multiput(1,4)(0.1,-0.1){30}{\circle*{0.01}}
  \multiput(1,1)(3,0){2}{\circle*{0.15}}
  \multiput(1,4)(3,0){2}{\circle*{0.15}}
  \put(0.75,0.25){$x_1$}
  \put(4,0.25){$x_2$}
  \put(0.75,4.5){$x'_1$}
  \put(4,4.5){$x'_2$}
  \put(2.25,0.5){$a_{12}$}
  \put(2.25,4.3){$a_{12}$}
  \put(0.25,2.5){$b_{11}$}
  \put(4.25,2.5){$b_{22}$}
  \put(0,3.9){$\frac{1}{2}a_{11}$}
  \put(0,0.9){$\frac{1}{2}a_{11}$}
  \put(4.25,3.9){$\frac{1}{2}a_{22}$}
  \put(4.25,0.9){$\frac{1}{2}a_{22}$}
  \put(2.25,3){$b_{12}$}
  \end{picture}
  \end{center}

\begin{figure}
 \begin{center}
 \epsfxsize=100mm
 \epsfysize=80mm
 \epsffile{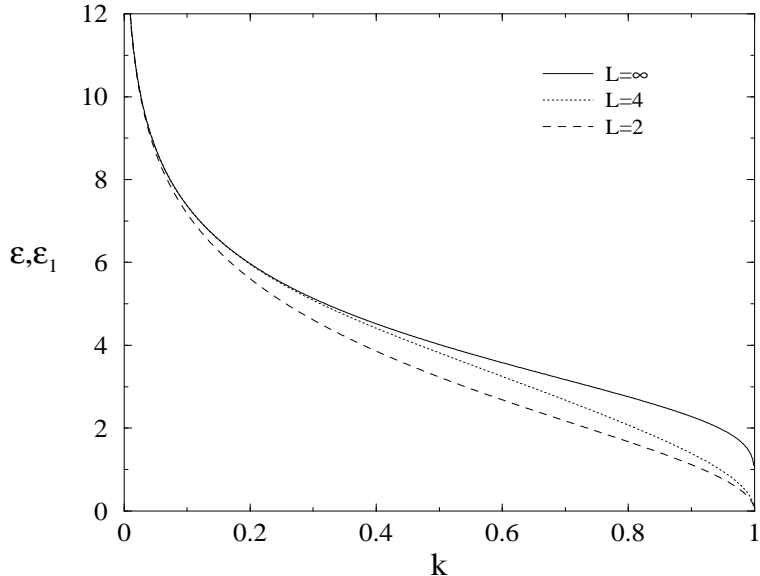}
 \caption{\label{fig4} Lowest eigenvalues in the density 
  matrix of a half-chain. Plotted are $\varepsilon$ for $L=2$ and
  $\varepsilon_1$ for $L=4$ and $L=\infty$. }
 \end{center}
 \end{figure}

  The cross-couplings, shown as dashed lines, can be eliminated by
  introducing new coodinates $y_i,y'_i$. After that, a sequence of
  transformations similar to those in Section \ref{sec:one} brings
  $\rho_h^{1/2}$ (and thus $\rho_h$) into diagonal form. Some of the
  details are given in Appendix~\ref{app:four}. The final result is
  that $\rho_h$ has also the form (\ref{eqn:calH}) where now
  \begin{equation} {\cal{H}}\;=\;\sum_{j=1}^2\,\varepsilon_{j}
  \beta_j^{\dagger} \beta_j \label{eqn:fH} \end{equation} describes
  {\it two} harmonic oscillators with energies $\varepsilon_1$ and
  $\varepsilon_2$. Thus one obtains a simple generalization of the
  case $L=2$. Also the variation of $\varepsilon_1$ with $k$ is very
  similar to that of $\varepsilon$ in Section \ref{sec:one}. This is
  shown in Fig.\ref{fig4}, where both quantities are plotted.  In
  particular, one finds that they coincide in the limit $k\rightarrow
  0$. The ratio $\varepsilon_2/\varepsilon_1$ equals $3$ for small k,
  drops to a minimum of $2.866$ for $k=0.34$ and then increases
  continuously, because $\varepsilon_2$, in contrast to
  $\varepsilon_1$, stays finite as $k\rightarrow 1$.  The shape of the
  spectrum, which depends on the ratio $\varepsilon_2/\varepsilon_1$,
  will be discussed in the next section.

   At this point one can already conjecture that the structure of
   $\rho_h$ remains the same also for larger $L$. A direct computation
   as above does not seen feasible, though. In the limit of large $L$,
   however, a different approach is possible. As in \cite{nish,peschk}
   one first relates $\rho_h$ to the partition function of a
   two-dimensional classical system, which is a massive Gaussian model
   in our case, in the form of an infinite strip of width $L$ with a
   perpendicular cut. This connection is discussed in more detail in
   Appendix \ref{app:gauss}.  One then expresses the partition
   function as the product of four corner transfer matrices. In the
   case where $L$ is much larger than the correlation length, one can
   use the thermodynamic limit of these CTMs and finds for $\rho_h$
   the form (\ref{eqn:calH}) with an operator $\cal{H}$ which is very
   similar to $H$ in (\ref{eqn:H}). The coefficients, however, are
   multiplied by additional site-dependent factors which increase
   linearly along the chain and reflect the corner geometry.  Up to a
   prefactor it is the operator given in (\ref{eqn:tran}) in Appendix
   \ref{app:gauss} and its diagonalization amounts to finding the
   normal modes of the corresponding vibrational problem.  From the
   results in \cite{pescht} one obtains \begin{equation} {\cal
   H}\;=\;\sum_{j\geq1} \,(2j-1)\,\varepsilon\,\beta_j^{\dagger}
   \,\beta_j \label{eqn:infH} \end{equation} with \begin{equation}
   \varepsilon\;=\;\pi\,\frac{I({k'})}{I(k)} \label{eqn:ellipt}
   \end{equation} where $I(k)$ is the complete elliptic integral of
   the first kind and ${k'} = \sqrt{1-k^2}$. Therefore ${\cal H}$
   describes an infinite set of harmonic oscillators with energies
   $\varepsilon_j=(2j-1)\varepsilon$ and is a straightforward
   extension of the results for small $L$.
   
   The parameter $\varepsilon\equiv\varepsilon_1 $ is also shown 
   in Fig.{\ref{fig4}}. For $k\rightarrow~0$,
   it has exactly the same expansion as $\varepsilon$ for $L=2$ and 
   $\varepsilon_1$ for $L=4$. For $k\rightarrow~1$, it vanishes only 
   logarithmically, i.e. more slowly than the quantities for finite $L$.

   One should note that the results (\ref{eqn:infH}),(\ref{eqn:ellipt}) 
   are formally the same as for the transverse Ising chain in the disordered 
   phase \cite{peschk}. The only difference is that there the operators 
   $\beta,\beta^{\dagger}$ are fermionic (so that $\beta^{\dagger}\beta=0,1$),
   whereas here they are bosonic. Such similarities can also be observed 
   in the row transfer matrices of the Gaussian and the Ising model, if one 
   uses the corresponding parametrizations \cite{sato}. The consequences for 
   the spectrum of $\rho_h$ are discussed below.

  \begin{figure}
  \begin{center}
  \epsfxsize=100mm
  \epsfysize=80mm
  \epsffile{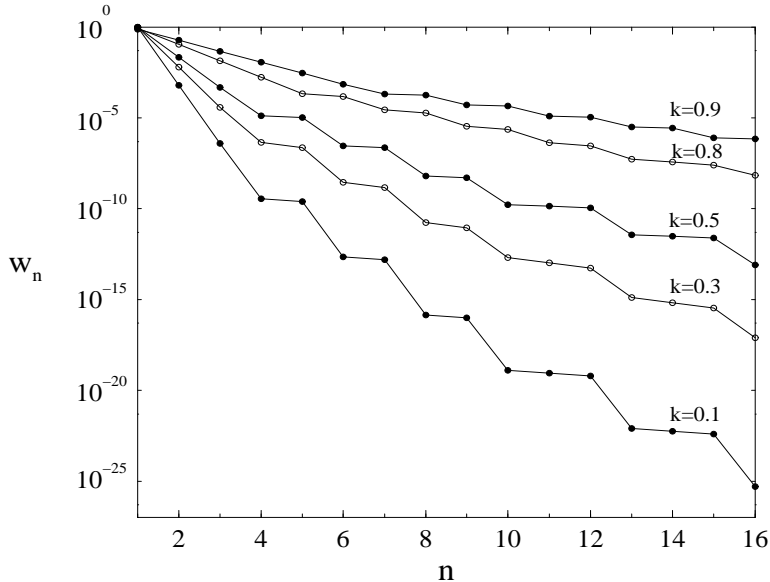}
  \caption{\label{fig5} Density-matrix spectrum for $L=4$
          and five different values of $k$.}
  \end{center}
  \end{figure} 
   
   \section{Spectra and Numerics} \label{sec:spectra}   
   In the following, we show the density-matrix spectra for
   half-chains and discuss some numerical aspects.
   In the figures, the eigenvalues $w_n$ of 
   $\rho_h$ are ordered according to magnitude 
   and plotted on a semilogarithmic scale.
   
   Figure \ref{fig5} shows spectra for $L=4$ and several values of
   $k$. These results were obtained by calculating the two energies
   $\varepsilon_1,\varepsilon_2$ numerically from the formulae in
   Appendix~\ref{app:four}.  Apart from the rapid decrease, one notes
   a clear ladder structure for the smallest three $k$'s.  It results
   from the relation $\varepsilon_2\cong~3\varepsilon_1$ which leads
   to the approximate degeneracies $(1,1,1,2,2,2,3)$ for the first
   seven levels. The steps for $k=0.3$ are less perfect, since
   $\varepsilon_2$ deviates more from $3\varepsilon_1$ in this case.
   For the two largest $k$'s, $\varepsilon_2\cong~4\varepsilon_1$ and
   $\varepsilon_2\cong~6\varepsilon_1$, so that the first step appears
   at these levels and the spectra look more stretched out.
     
\begin{figure}[ht]
  \begin{center}
  \epsfxsize=70mm
  \epsfysize=90mm
  \epsffile{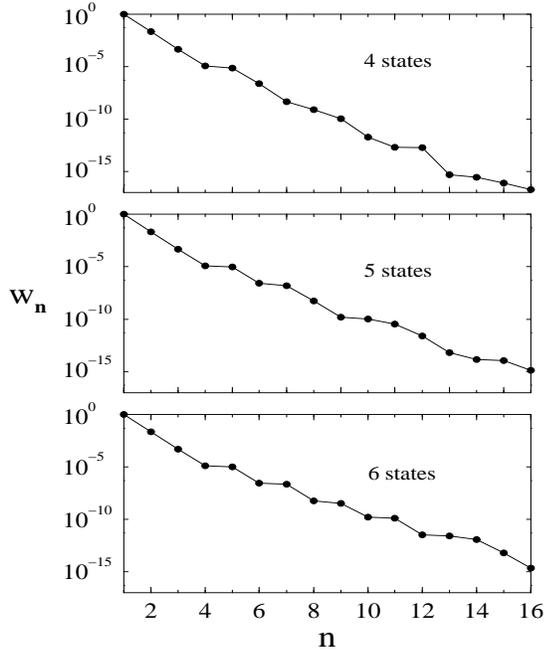}
  \caption{\label{fig6} Density-matrix spectrum for
    $L=4$ and $k=0.5$, calculated with different numbers 
    of oscillator states.}
  \end{center}
  \end{figure}

   It is interesting to see, how these results are recovered in a
   numerical treatment using a truncated Hilbert space.  If one works
   with the eigenstates of $\rho_1$, a small number $(5-7)$ is
   sufficient for not too large $k$.  For example, if $k=0.5$ and one
   chooses the same $r$ states (with some average $\varepsilon$-value)
   for all four sites, the error in the ground-state energy $E_0/L$ is
   of the order $10^{-r}$. The spectra which then result are shown in
   Fig.\ref{fig6} for three values of $r$.  The first $w_n$ are always
   quite accurate, but there are characteristic differences for the
   following ones, which are connected with the number of steps,
   i.e. with the degeneracies. One can see that if $r$ states are
   kept, the pattern is correct for the first $r$ levels (counted from
   the top). At the next level, the state with energy $r\varepsilon_1$
   is missing and the corresponding step is absent. Thus there is a
   certain correspondence between the states in the local basis and in
   the density matrix.  For smaller $w_n$, however, the situation is
   less clear, and the spectrum finally becomes irregular. The tails
   of the approximate spectra always lie below the exact one.
   
\begin{figure}[ht]
  \begin{center}
  \epsfxsize=100mm
  \epsffile{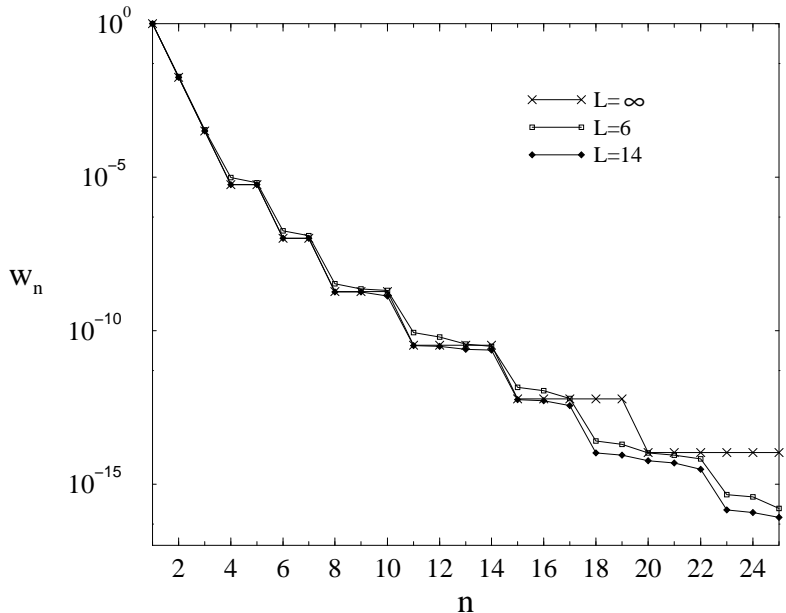}
  \caption{\label{fig7} Density-matrix spectrum for
         $k=0.5$ and two sizes $L$, calculated
        with DMRG using $7$ states and $m=7$.
        Also shown is the analytical result for 
        $L\rightarrow~\infty$. }
  \end{center}
  \end{figure}

  In order to obtain results also for $L>4$, we have carried 
  out DMRG calculations, using $7$ states at each site, 
  with an $\varepsilon$ corresponding to $L=30$. With $m=7$
  kept states per block, the error in $E_0/L$ was about  
  $3\cdot~10^{-7}$ for $k=0.5$. Fig.\ref{fig7} 
  shows the resulting spectra for $L=6$ and $L=14$,
  together with the thermodynamic limit according to
  (\ref{eqn:infH}),(\ref{eqn:ellipt}). One notes, 
  that the spectra for the two $L$'s are similar,
  though not identical. Compared to $L=4$, 
  the degeneracies have changed to $(1,1,1,2,2,3,4)$.
  The latter two result from a third energy 
  $\varepsilon_3\cong~5\varepsilon_1$, which first 
  appears for $L=6$. This shows that, indeed, the number of 
  oscillators in $\rho_h$ is equal to the size $L/2$
  of the half-chain. One also sees that for $L=14$ 
  the first two steps have become perfect,
  so that $\varepsilon_2=3\varepsilon_1$ 
  as for the infinite system. Up to some small deviations,
  this also holds for the next two steps. Only for the remaining 
  levels $8$ and $9$, the degeneracies are not correct. 
  This is the same effect as found above for $L=4$.

  For $L=14$, the $\varepsilon_j$ are also numerically very close
  to the large-$L$ limit. For example, $\varepsilon_1$ agrees
  with the exact result $4.0189$ up to three decimal places.
  This can be understood from the short correlation length
  $\xi\sim~3$ for $k=0.5$ which makes size-effects small. 
  Finally, we want to mention that, in the thermodynamic
  limit, the multiplicities are just one-half of those 
  found in the fermionic case for the ordered phase 
  where $\varepsilon_j=2\varepsilon j$ \cite{peschk}. 
  This is because the number $P_j$ of partitions
  without repetition is the same as that of the odd 
  integers with repetition, $P_j=P'_{2j-1}$. 
  Therefore the degeneracies for the bosonic case 
  are not as large as one might expect at first. 

   \section{Conclusion} \label{sec:concl}
    We have investigated a bosonic system, where the ground-state
    density matrices can be determined explicitely 
    in various cases. It turned out that they are exponentials of oscillator
    Hamiltonians, so that all results are quite transparent. 
    The spectra have exponential character and 
    the eigenfunctions are oscillator states. For the 
    single-site density matrix, these states are related 
    to those of the chain oscillators by squeezing.
    For the half-chain density matrix, they are connected with 
    certain normal modes concentrated near the middle of the system.
    The thermodynamic limit was obtained in the same way 
    as for the integrable spin chains treated previously,
    and the spectra are very similar to those found there.
    By counting the degeneracies, one would arrive at 
    formulae as given in \cite{OHA}.

   Taking all this together, the chain treated here may serve as a 
   standard example where one can see the features of 
   the density matrices in detail. In this context, it would
   still be interesting to determine the half-chain density matrix
   for arbitrary sizes, in particular at the critical point,
   where the vibrational spectrum becomes acoustic.
   This case has already been studied by DMRG \cite{caron} but, 
   as for the critical spin models, the density-matrix spectra 
   have yet to be explained.
   Another question is, whether the model of coupled oscillators,
   for which the ground state is known explicitely,
   could be used to study density matrices in higher dimensions.
   For the DMRG method, it would be quite important to know
   if the spectral properties change in this case.
  

  \noindent{\large\bf Acknowledgements} 
  
  \vspace{0.2cm}
  We thank M.~Kaulke, E.~Jeckelmann, G.~Babudjan,
  A.~Pelster and H.~Kleinert for
  discussions. M.C.~Chung acknowledges the support of Deutscher 
  Akademischer Austauschdienst (DAAD).
  \section*{Appendices}
  \appendix
  \section{Four Oscillators} \label{app:four}
   In order to diagonalize $\rho_h^{1/2}$ in (\ref{eqn:fourd}) one proceeds 
   as follows. First, new coordinates are introduced by a rotation $(x_1,x_2)
   \rightarrow (y_1,y_2)$ with angle $\varphi$ and analogously for the primed
   quantities. This leads to new quadratic forms in the exponent, with
   coefficients $\widehat{a}_{ij}$ and $\widehat{b}_{ij}$. Choosing 
   $\tan{2\varphi} = 2b_{12}/(b_{11}-b_{22})$, the cross-term 
   $\widehat{b}_{12}$ becomes zero. One then considers the factors
   \begin{equation}
    \exp{(-1/2 \,\widehat{a}_{ii}\, y_i^2\,)}\,
    \exp{(-\widehat{b}_{ii} \,y_iy'_i\,)}\,
    \exp{(-1/2\,\widehat{a}_{ii}\, {y'_i}^2\,)}~~~,i=1,2
   \end{equation}
   which contain only $(y_i,{y'}_i)$. These can be transformed as in Section
   \ref{sec:one} and one obtains exponentials of harmonic oscillators with 
   energies 
   \begin{equation}
    \nu_i\; = \;2\sinh^{-1} (\Omega_i/2) 
   \end{equation}      
   where
   \begin{equation}
     \Omega_i/2 \;=\; \sqrt{(\widehat{a}_{ii}+\widehat{b}_{ii})/
     (-2\,\widehat{b}_{ii})}
   \end{equation}
   In terms of the new coordinates $z_i$ one then has
   \begin{equation}
    \rho_h^{1/2} \;=\; C\cdot\,\exp{(-\mu z_1z_2)}\,\exp{(-\sum_{i}
    (-\frac{1}{2} \frac{\partial^2}{\partial z_i^2}+\frac{1}{2}
    \nu_i^2z_i^2\,)\;)}\, \exp{(-\mu z_1z_2)} \label{eqn:z}
   \end{equation}
   Here $z_i =y_i/\lambda_i,\,\mu = \widehat{a}_{12}\,\lambda_1\lambda_2$
   and the $\lambda_i$ are given by
   \begin{equation}
    \lambda_i \;=\; \left(\frac{\nu_i}{-\widehat{b}_{ii} \Omega_i}\right)^{1/2}
    \, \left(1+\frac{\Omega_i^2}{4}\right)^{-1/4} 
   \end{equation}
   In the final step, one expresses (\ref{eqn:z}) in terms of bosonic operators
   $\alpha_i,\alpha_i^{\dagger}$ and considers 
    Heisenberg-like operators $\rho_h^{1/2} \alpha_i \rho_h^{-1/2}$ 
   which are found to be linear combinations of 
   the $\alpha_i,\alpha_i^{\dagger}$. Therefore, a transformation as in the 
   analogous fermionic case \cite{lieb}
   \begin{equation}
    \beta_{j} \;=\; \sum_i\, (g_{ji} \alpha_i+ h_{ji} \alpha_i^{\dagger})
    \end{equation}
   brings $\rho_h^{1/2}$ into the form (\ref{eqn:calH}),(\ref{eqn:fH})
   with energies $\varepsilon_j /2$. These energies follow from a simple 
   quadratic equation, namely
   \begin{equation}
     \cosh\frac{\varepsilon_j}{2} \;=\; \frac{1}{2}(c_1+c_2) \pm 
     \sqrt{\frac{1}{4}(c_1-c_2)^2+4\rho^2s_1s_2}
     \label{eqn:fenergy}
   \end{equation}
    where  $c_i =\cosh\nu_i, s_i=\sinh\nu_i \,{\mbox{and}}\, \rho= \mu/2
    \sqrt{\nu_1\nu_2}.$

  These quantities have to be evaluated, starting from the initial  constants
   $a_{ij}$ and $b_{ij}$, which are simple analytic expressions involving the 
    four eigenfrequencies of the chain. It turns out that, for small values
    of $k$, the $\rho$-term in (\ref{eqn:fenergy}) is unimportant, which leads
    to $\varepsilon_j \cong 2\nu_j$ and $\varepsilon_2 \cong 3\varepsilon_1$.
    The ratio $\varepsilon_2/\varepsilon_1$ thus has the same value as in the 
    thermodynamic limit. Moreover,
     $\varepsilon_1$ has the same asymptotic form, 
    $\varepsilon_1 \cong 2 \ln(4/k)$, as for $L=2$ and $L \rightarrow \infty$.
    This can be attributed to the short correlation length which suppresses
    size  effects in this limit.
    
  \section{Relation to the Gaussian Model}\label{app:gauss}
   The Hamiltonian $H$ in (\ref{eqn:H}) has a close relation to the transfer
   matrix of a two-dimensional Gaussian model (GM). The connection is the same 
   as between the transverse Ising chain and the two-dimensional Ising model
   \cite{peschk}. Consider a lattice with variables ${x\; (-\infty < x <
    \infty)}$ at each site, a nearest-neighbour coupling energy 
   $\frac{1}{2}\,K\, (x-{x'})^2$ and an on-site energy $\frac{1}{2}
   \,\Delta\,x^2$, all in units of $k_B T$. If 
    the lattice is oriented diagonally, the appropriate 
    transfer matrix $T$ involves the piece shown in the figure below.

  \begin{center}
  \unitlength1cm
  \begin{picture}(8,4)
  \hspace*{-1cm}
  \multiput(1.0,3.0)(2,0){4}{\line(1,-1){2}}
  \multiput(1.0,1.0)(2,0){4}{\line(1,1){2}}
  \put(0.5,1.55){\line(1,-1){0.5}}
  \put(9,3){\line(1,-1){0.5}}
  \put(0.5,2.5){\line(1,1){0.5}}
  \put(9,1){\line(1,1){0.5}}
  \multiput(1,1)(2,0){5}{\circle*{0.2}}
  \multiput(2,2)(2,0){4}{\circle*{0.2}}
  \multiput(1,3)(2,0){5}{\circle*{0.2}}
  \put(2.75,0.3){$x_{i-1}$}
  \put(4.75,0.3){$x_{i}$}
  \put(6.75,0.3){$x_{i+1}$}
  \put(2.75,3.3){${x'}_{i-1}$}
  \put(4.75,3.3){${x'}_{i}$}
  \put(6.75,3.3){${x'}_{i+1}$}
  \end{picture}
  \end{center}

  One can then verify by a simple direct 
  calculation (using two interpenetrating lattices)
   that, with periodic boundary conditions,
 \begin{equation}
  \left[H\, ,T\,\right] \:=\: 0 \label{eqn:commun}
 \end{equation} 
 provided that $k\,=\,K^2$ and $\omega_0^2\,=\,\Delta(\Delta+4K)$. In this 
 case, $T$ and $H$ have common eigenfunctions and $\Psi$ in (\ref{eqn:psi})
 gives the maximal eigenvalue for $T$. This allows one to obtain $\Psi$
 and also $\rho_h$ from the partition function of a two-dimensional system
 \cite{peschk,nish}. If the GM has open boundaries, one has to modify
   $H$ at the end, so as to preserve (\ref{eqn:commun}). However, for a system 
  with $L \gg \xi$, where $\xi$ is the correlation length
  given by $\xi=2/\ln{(1/k)}$, this effect 
  is not important and can be neglected.

  An alternative approach is to consider a Gaussian model with anisotropic
   couplings for periodic boundary conditions, to show that the ${T'}$s for
   different anisotropies commute and to realize that a proper derivative 
   leads to $H$ \cite{babu}. To do this, one uses an elliptic parametrization,
   so that the two couplings are, for example,
  \begin{equation}
   K_1\,=\,-i\,/\,\sn(iu,k)~~~;~K_2\,=\,i\,k\,\sn(iu,k)
  \end{equation}  
  with the Jacobi function $sn$ of modulus $k$. This parameter also 
  determines  the on-site energy $\Delta$ and thus the distance to the 
  critical point $\Delta\,=\,0$, as well as the correlation length.
   The parameter $u$, on the other hand, specifies the 
   ratio $K_1/K_2$. It varies between $0$ and $I({k'})$, where
   $I$ is the complete elliptic integral of the first kind and
  ${k'}\,=\,\sqrt{1-k^2}$. The isotropic case corresponds to $u\,=\,I({k'})/2$.
  (Our notation differs slightly from that in \cite{babu}. We have interchanged
   $k \leftrightarrow {k'}$, written $u$ instead of $\lambda\theta$, used 
   $x\,=\,\sqrt{\lambda}\phi$ for the Gaussian variables and have set 
   $\alpha\,=\,-1$.) The derivative $(\partial \ln T/ \partial u)$ then leads
   again to (\ref{eqn:H}) with $\omega_0\,=\,(1-k)$, which is the reason for 
   choosing this parametrization in $H$.

  As discussed in \cite{peschk}, the density matrix $\rho_h$ for half of 
  the system is, for $L \gg \xi$ and up to a prefactor,
  \begin{equation}
    \rho_h\;=\;ABCD
  \end{equation} 
  where $A,B,C,D$ are the corner transfer matrices for the four infinite 
  quadrants of the two-dimensional system. Due to the integrability of
  the Gaussian model, i.e. the Yang-Baxter equations, these have 
  the exponential form   
   \begin{equation}
     A\;=\; e^{-u\,{\cal{H}}_{\rm CTM}}
   \end{equation}
   and similarly for $B,C,D$, with ${\cal{H}_{\rm CTM}}$ given by
    \begin{equation}
    {\cal{H}}_{CTM} = \sum_{n \geq 1}\left\{-\frac{1}{2}(2n-1)
     \frac{\partial^2}{\partial x_n^2}+
     \frac{1}{2}(2n-1)(1-k)^2 x_n^2+\frac{1}{2}
     2nk(x_{n+1}-x_n)^2\right\}  \label{eqn:tran}
    \end{equation}
    This operator was studied
   in \cite{pescht} in connection with the Hamiltonian limit $u \rightarrow 0$
   of $A$, where one can determine its form simply by inspection. 
   It is associated with a corner of Ramond type, i.e. without
   a site at the tip. In terms of vibrations,
   it describes a system of coupled oscillators, 
   where the spring constants and inverse masses increase 
   along the chain. It can
   be diagonalized with the help of Carlitz polynomials
   and then becomes the sum
   of harmonic oscillators with eigenvalues $(2j-1)\pi /2 I({k'})$.
   For $\rho_h$ one needs $ABCD$, or $A^4$ if one has an isotropic model. In 
   either case this gives a factor $2I({k'})$, so that the energy becomes 
   $\varepsilon_j=(2j-1)\pi I({k'})/I(k)$ and one arrives at the result
   (\ref{eqn:infH}),(\ref{eqn:ellipt}).

 \end{document}